\documentclass[twocolumn,preprint]{aastex631}

\usepackage{amssymb}
\usepackage{amsthm}

\usepackage{amsmath}
\usepackage{mathrsfs}
\usepackage[mathscr]{euscript}

\usepackage{color,soul}

\begin{document}

\title{Motivating Emissions from Positive Energy Warp Bubbles}

\author[0000-0001-7807-4825]{Erik W. Lentz}
\affiliation{Pacific Northwest National Laboratory \\ 
902 Battelle Blvd. Richland, WA 99352, USA}

\author[0000-0002-0766-4163]{Ryan C. Felton}
\affiliation{NASA Ames Research Center, Mountain View, CA 94035, USA}



\begin{abstract}

Recent research has proposed that advanced propulsion mechanisms such as warp drives are more physically feasible than previously thought, using positive energy sources potentially sourced by known classical physics. Motivated by this, we hypothesize that an advanced inter-planetary or interstellar civilization using warp drives at sub-luminal or super-luminal speeds will broadcast detectable emissions of their travels. These technosignatures would be of significant astronomical, physical, and technological interest. This paper seeks to motivate signatures from warp drive emissions due to intrinsic and extrinsic processes across several messenger types (electromagnetic, particle, and gravitational) and propose a research program to simulate such emissions in sufficient detail to search for their signatures through coordinated analyses across multiple observatories.

\end{abstract}



\section{Introduction/Motivation}
\label{sec:introduction}

Spacecraft technosignatures are one of many potential signatures of extraterrestrial intelligence (ETI) \citep{HAQQMISRA2022194}. Unlike quasi-static planetary-based or stellar-based engineered structures tied to well-defined marker objects~\citep{NASATechnosigWorkshop,Sheikh2021Technosignatures,Romanovskaya_2022}, spacecraft by their design are made to quickly span increasingly ambitious inter-planetary, inter-stellar, or inter-galactic distances. Nonetheless, an ETI-driven spacecraft may produce unique signatures across its path through interaction with its immediate surroundings, such as the interstellar medium, or intrinsically as a byproduct of the craft's powerplant, propulsion mechanism, control system, etc., or some interaction thereof~\citep{Zubrin1994,Romanovskaya_2022}. These signatures may highlight points of interest, yet to be detected exoplanets, or other points of congregation, providing focal points for the astrobiology community's research interests. 

Current considerations of ETI craft mobility are rather limited, focusing on inertial concepts of propulsion such as rocketry or light sails to move a craft and use costly propellants or other means of momentum transfer to boost the craft towards its destination, relegating it to sub-luminal speeds~\citep{Zubrin1994,NASATechnosigWorkshop,Sheikh2021Technosignatures,HAQQMISRA2022194,Romanovskaya_2022}. Such mechanisms are highly resource intensive and come with long travel times that may be considered wasteful by an ETI. There may be a point where even the bonds of special relativity are undesirable, motivating the invention of a means of transport capable of effectively super-luminal rates. Such means of super-luminal travel may involve concepts of fundamental physics and methods of advanced engineering as yet to us unconfirmed or unknown and would naturally be of intense interest to both the physics and engineering/technology communities, though we do not consider it useful to speculate on these concepts at this time.

Physics research does point to one possibility using known classical physics and Einstein's general theory of relativity to generate compact regions of curvature in space-times capable of transporting observers up to and potentially beyond the speed of light, commonly known as warp drives \citep{Lentz2020,Bobrick2021,Fell2021,huey2023membrane,Fuchs_2024}. We hypothesize that a sufficiently advanced civilization having obtained the capability to use warp drives or similar as a practical transportation device may be observable when traveling between points of interest of that civilization. We therefore propose to theoretically model and search for emissions  from craft moving at sub- and super-luminal speeds between stellar systems in our galaxy. The craft's technosignatures are expected to be numerous and varied, resulting from both intrinsic operations and interactions with its environment, extending across several messengers (electromagnetic, particle, and gravitational), and occurring during each stage of travel (acceleration, coast, and deceleration).

The remainder of this paper seeks to motivate emissions from warp drives by computing preliminary signal estimates and outlining a research plan resulting in searches for more complete and realistic signals. 
Section~\ref{sec:warpreview} presents a brief overview of the state of warp drive research in the literature, followed by recommendations for next steps leading to modeling of a drive's emissions. 
Section~\ref{sec:signatures} outlines categories of emissions from a warp drive and the resulting signatures incident on an observing instrument, providing several examples of estimated signatures carried via electromagnetic, gravitational, and massive particle messengers. 
Section~\ref{sec:program} outlines a research program to model physical warp drives, simulate synthetic emissions over the course of a realistic journey as they would appear in a telescope or other instrument, and mount searches starting close to the Earth and working further out into the Galaxy. 
Section~\ref{sec:closing} closes the paper with a summary and next steps to initiating the research program.

\section{A Brief Primer on Warp Drive Research}
\label{sec:warpreview}

The topic of super-luminal travel in general, and warp drives in particular, has been a topic of fascination for scientists and the public alike for over a century~\citep{keplersomnium_Rosen1967,wells1901first,smith2011skylark,campbellislands}. Warp drives as a topic of scientific study became widely recognized three decades ago with a collection of papers initiated by \citet{Alcubierre1994} as a means of using the intrinsic dynamics of space--time as governed by Einstein's general relativity (GR) to transport a spacecraft at arbitrary speed as opposed to purely inertial mechanisms. The first two and a half decades that followed added numerous analyses of and improvements to the Alcubierre solution as well as multiple novel space--times  \citep{Everett1996,Pfenning1997,Hiscock1997,Krasnikov1998,Olum1998,VanDenBroeck1999,Millis1999,Visser2000,Loup2001,Natario2002,Gauthier2002,Lobo2003,Lobo2004,Lobo2007,Finazzi2009,McMonigal_2012}, 
some of which used approaches to gravity outside that of typical GR in three space and one time dimensions \citep{Obousy2008,White2013,DeBenedictis_2018}. We will confine our treatment of space--time dynamics to GR in 3+1 dimensions unless otherwise stated.

For reference, we here subscribe to the description of a warp drive space--time as a nested bubble-like configuration with an innermost extended yet compact region that is (nearly) flat such that it is safe for passengers, enclosed by another compact region that is arbitrarily curved, and lastly surrounded by an asymptotically flat or hyperbolic vacuum region. This definition aligns well with the prescription given by \citet{Bobrick2021} and \citet{Fuchs_2024}. The space--time metric describing the warp bubble geometries is most often decomposed according to the Arnowitt-Deser-Misner (ADM) formalism \citep{ADM}. This decomposes the infinitesimal path length as
\begin{align}
    ds^2 &= g_{\alpha \beta} dx^{\alpha} dx^{\beta} \\ \nonumber
    &= -\left(N^2-N^i N_i \right) dt^2 - 2 N_i dx^i dt + h_{ij} dx^i dx^j,
\end{align}
where the time coordinate $t$ stratifies space--time into space-like hypersurfaces, the space metric components $h_{ij}$ evaluated at $t$ provide the intrinsic geometry of that hypersurface, and the similarly-evaluated shift vector components $N^i$ at $t$ provide the coordinate three-velocity of the hypersurface's normal.  The time-like unit normal one-form is therefore proportional to the coordinate time element $\mathbf{n}^* = N dt$. Einstein summation notation is used throughout this paper, with Greek indices running over space--time components and Latin indices over space components. The lowering of Latin indices is performed using the hypersurface metric $h$ unless otherwise stated. Natural units $G=c=1$ are used for simplicity of form. The geometries researched in this period were largely of the ``Natario class''~\citep{Natario2002} where the lapse function $N$ is set to unity and the hypersurface metric is set to be flat. $h_{ij} = \delta_{ij}$ in Cartesian coordinates where $\delta_{(\cdot \cdot)}$ is the Kronecker delta function. The non-flat geometry is encoded in the three-component shift vector, $N_i$.

There have been many justified critiques regarding the feasibility of these early space--times, the most prominent noting that the geometries must largely be sourced from a form of negative energy density, which has no known macroscopic source in particle physics~\citep{Olum1998,Visser2000,Lobo2003,Santiago2021a}. Other concerns include the immense (absolute value) energy requirements to create a bubble, the difficulty associated with constructing a bubble from a nearly flat space--time plus source stress-energy-momentum (SEM) up to the super-luminal phase, where the transported central observers are expected to become surrounded by a horizon, and the difficulties of driving the super-luminal phase back to the nearly flat space--time. Some significant progress was made to reduce the energy magnitude requirements~\citep{VanDenBroeck1999,Loup2001,Krasnikov2003,Obousy2008,White2013}, but the remaining challenges persisted to the beginning of this decade.

The challenge of physicality of a warp drive's sourcing media has been revisited in the last several years \citep{Lentz2020,Bobrick2021,Fell2021,huey2023membrane,Fuchs_2024} with a focus on producing geometries that require only positive energy sources and eventually satisfying the set of guidelines known as the energy conditions (null, weak, strong, and dominant)~\citep{MTW}. A warp drive satisfying all four energy conditions~\cite{Fuchs_2024} would be far more feasible in a laboratory setting, though the high energy requirements and construction mechanism of a bubble remain as challenges. While human efforts to design let alone construct a complete warp bubble in the lab are yet to be seen, perhaps a sufficiently advanced ETI has resolved these challenges and made warp drives or similar into a practical means of transport. The next section considers emissions that may be produced during the operation of a warp drive. 

\section{Preliminary Estimates of Techno-signatures}
\label{sec:signatures}

There already exists some discussion of techno-signature emissions from warp drives and other crafts undergoing extreme speeds and/or momentum transfers in the literature~\citep{McMonigal_2012,Sellers2022}. These treatments respectively consider emissions from interaction of the Alcubierre metric with an external stress-energy medium such as the ISM and gravitational wave (GW) emissions intrinsic to the acceleration of a massive craft. This section aims to outline a broader set of emissions caused by warp drives and provide some preliminary estimates of their form and intensity. Here the emission types are organized into two categories: intrinsic emissions originating from internal processes of the bubble, and extrinsic emissions which involve interaction of the bubble with its environment and parameters both external and internal to the bubble. The range of detectability of emissions in each messenger type is also estimated to help guide potential searches.

Bubbles are comprised of stress-energy-momentum and space--time curvature coupled via Einstein's equation
\begin{equation}
    \mathbf{G} = 8 \pi \mathbf{T}
\end{equation}
where $\mathbf{G} = \mathbf{R} - 1/2 \mathbf{g} R$ is the Einstein two-tensor written in terms of the Ricci curvature tensor which itself can be considered as a function of the space--time metric and its first and second derivatives, and $\mathbf{T}$ is the SEM tensor of the physical fields. Emissions are any measurable excitations that propagate away from the bubble. They may take on the form of but are not limited to electromagnetic, gravitational, or massive particle radiation, and may transition between species as the local physics allows.

\begin{figure}
\begin{center}
\includegraphics[width=\columnwidth]{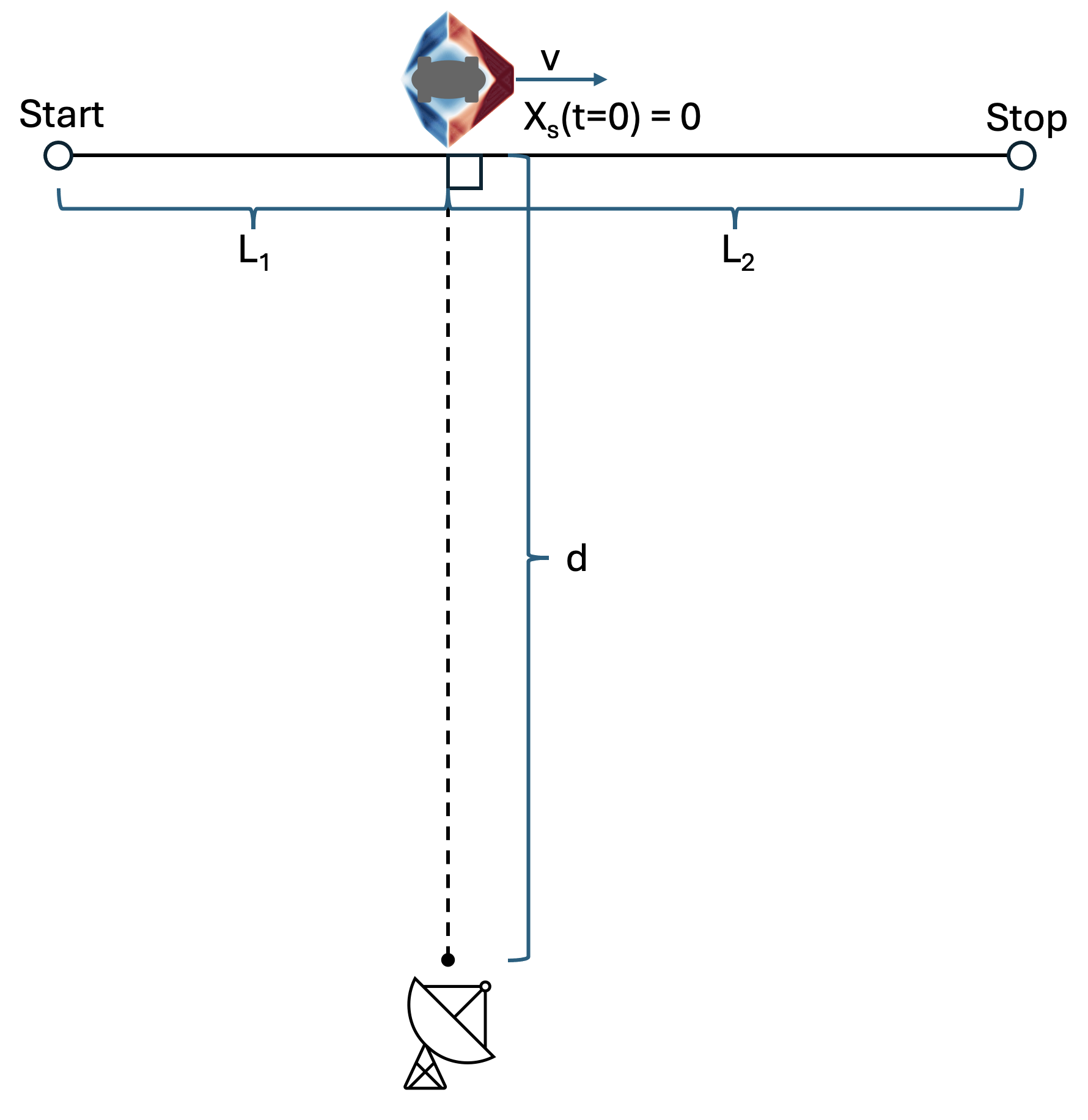}
\caption{Illustration of the bubble-observer system. The space--time well outside of the bubble is assumed to be effectively flat and will be parameterized using the rest frame of the observatory. The time of closest approach of the bubble to the observer is parameterized as $t=0$ and its distance $|\Delta x (t=0)| = d$. The craft's starting point along its trajectory is $L_1$ and its endpoint at $L_2>L_1$ relative to the point of closest approach. }
\label{fig:diagram}
\end{center}
\end{figure}

Let us prescribe a simple bubble-observer system from which to measure emissions, Fig.~\ref{fig:diagram}. The region of space--time away from the warp bubble will be taken as nearly flat meaning that the trajectories of emissions will be unperturbed by the space--time geometry once outside of the bubble's proximity, say outside a given distance from the bubble centroid $r>R_b$. The observation frame will be taken as inertial, though it is straightforward to correct for a terrestrial or nearby space-based observation platform in this limit. The observation frame's parameterization of time will be used to clock the emissions once they have entered the near-flat region outside the bubble. The spatial trajectory of the bubble centroid relative to the observation point will be parameterized in the observer time parameter as $\vec{r}_s (t) = \hat{x} x_s(t)  + \hat{y} d $ with the point of closest approach occurring at $t=0$, implying $x_s(0) = 0$. Emissions broadcast at a particular time $t_E$ and traveling at speed $v_E$ relative to the observation point will reach the observation point if moving undisturbed in a straight line at
\begin{equation}
    t_O = t_E + \frac{v_E}{c} \sqrt{d^2 + x_s(t_E)^2}. %v_s^2 t_E^2}.
\end{equation}

\begin{figure}
\begin{center}
\includegraphics[width=\columnwidth]{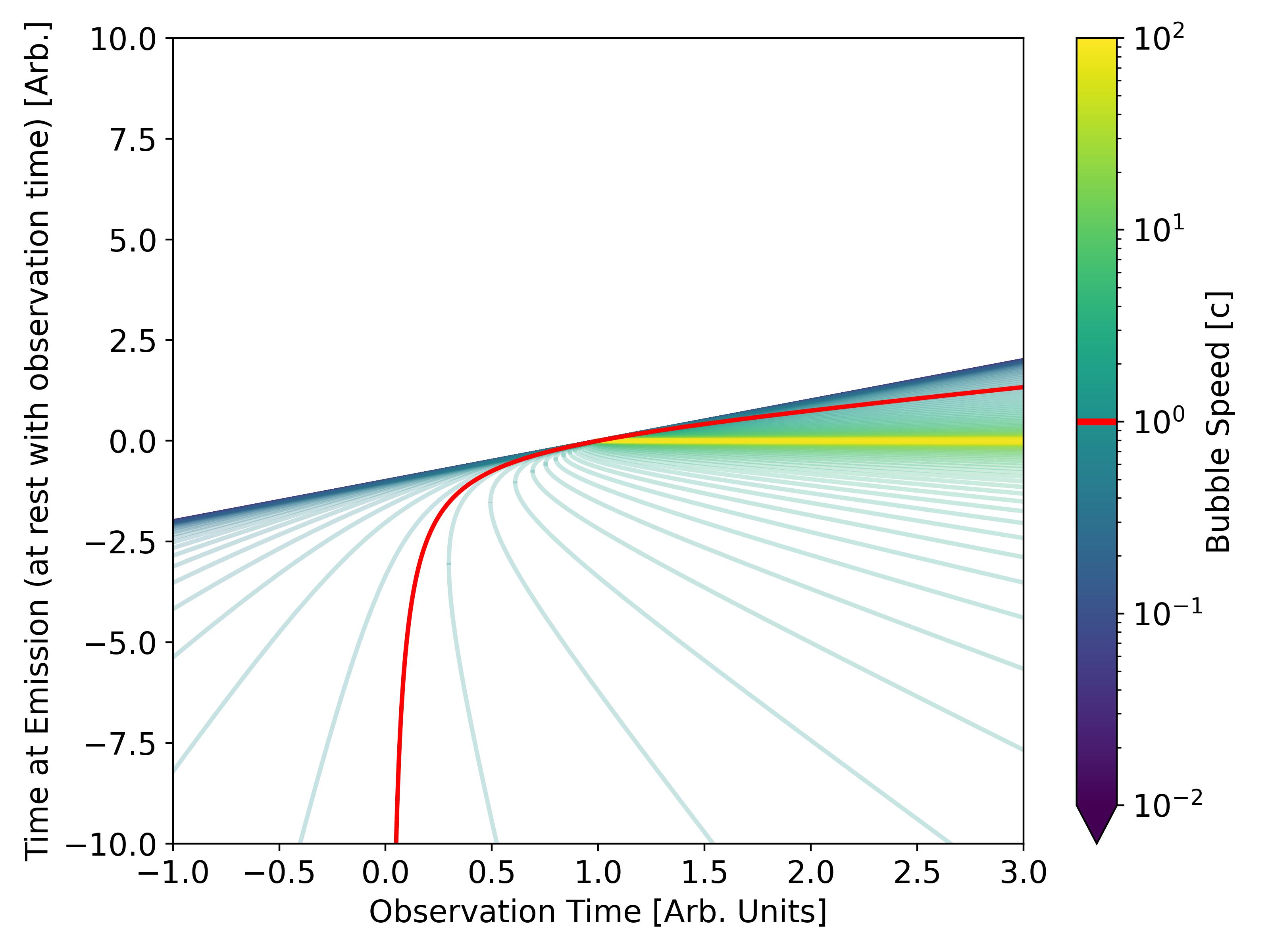}
\caption{Emission times viewed along an observer's timeline for light-like emissions from bubbles moving with prescribed constant speed with distance of closest approach set to $d=1$. The left-most point of each constant-speed path is referred to as the time of first light for the observatory. The light-speed craft is highlighted in red.}
\label{fig:times}
\end{center}
\end{figure}

\begin{figure}
\begin{center}
\includegraphics[width=\columnwidth]{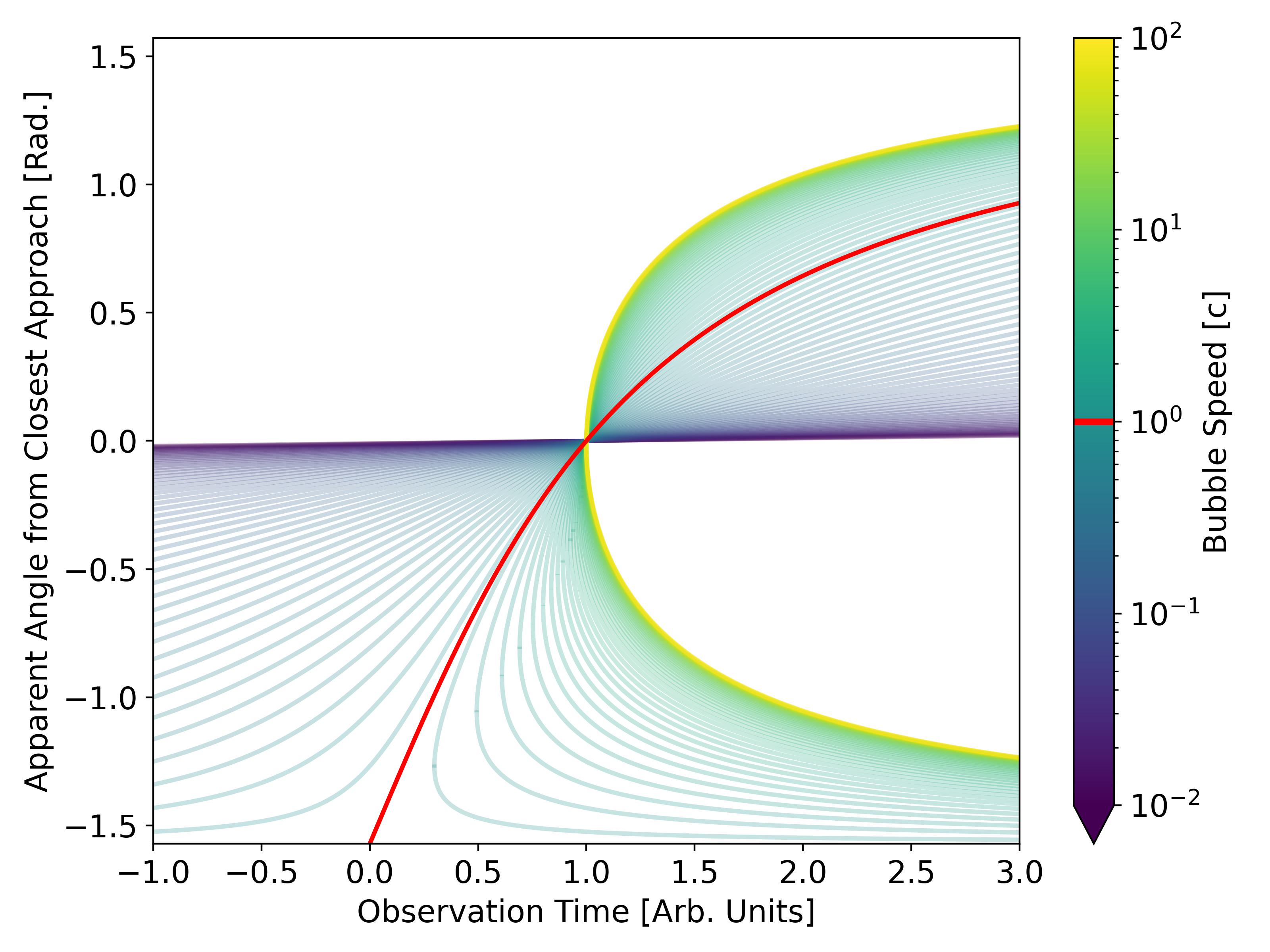}
\caption{Apparent emission angles viewed along an observer's timeline for light-like emissions from bubbles moving with prescribed constant speed with distance of closest approach set to $d=1$. The left-most point of each constant-speed path is referred to as the angle of first light for the observatory. The light-speed craft is highlighted in red.}
\label{fig:angles}
\end{center}
\end{figure}

All of this appears rather straightforward and even uninteresting until one considers the order of emissions as seen by the observer. For simplicity of this demonstration we take the bubble to move with constant speed $x_s(t) = v_s t$, also  $L_1, L_2 >0$, $L_1, L_2 \gg d$, and consider only light-like moving emissions, see Fig.~\ref{fig:times} and Fig.~\ref{fig:angles}. To correct for finite start and stop points of a bubble, the emission times ($t_E$) are truncated to $L_1/v_s < t_E < L_2/v_s$. Further, emission speeds lower than light speed imply an affine re-scaling of the observations time axis. For speeds $v_s >v_E$, the angular distribution of the observed signal will evolve bi-modally as the observatory records emissions that occurred at two different times on either side of the time/angle of first light. One signal moves in the apparent direction of travel of the bubble and presents emissions occurring forward in time while the other mode moves opposite and presents emissions in a reversed order. This second mode is purely due to the craft's motion which exceeds the emission's motion once separated from the bubble. It should be noted that not all finite bubble trajectories will produce bi-modal emissions at $v>v_E$, only those passing through the extremal emission time $t_E = -d/v_s|\sqrt{v_s^2 v_E^2 - 1}|$. The next subsection discusses opportunities for intrinsic emissions and presents several simplified example forms. 

\subsection{Intrinsic Emissions}
\label{sec:intrinsic}

We now consider emissions due to intrinsic processes of a `bare' bubble in an otherwise empty (flat) environment. Emissions from this bare bubble are expected during all phases of a craft's warp-enabled journey: acceleration, coasting, and deceleration. What we call acceleration consists of the formation of the warp bubble and apparent change in speed of the craft as viewed from an inertial frame in the asymptotically flat space--time outside the bubble. The coast phase has the bubble centroid moving with effectively constant velocity. The deceleration phase has the bubble weaken and diffuse and the craft change velocity, likely to match the destination's reference frame. While we may speculate the coast phase will last the majority of a journey, the acceleration and deceleration phases are expected to be more dynamic and prone to strong, complex, and bursty radiative losses. Tracking the generation of emissions throughout a warp bubble's lifecycle will ultimately require detailed numerical models evolving configurations of SEM and curvature practical for an ETI warp drive, which we do not have at this time. In Section~\ref{sec:program} we will outline first steps towards building these whole-journey models, but for the remainder of this section we estimate signals during the coast phase.

\subsubsection{Gravitational Waves}

Fluctuations in the space--time metric are capable of propagating across space--time in the far field in the form of waves of strain in the separation between nearby free-falling objects. We use the weak field limit for simplicity to linearize the Einstein equation in perturbations over the flat Minkowski metric $\eta$, expressed in components as
\begin{equation}
    g_{\mu \nu} = \eta_{\mu \nu} + h_{\mu \nu},
\end{equation}
despite the expectation that the linear limit is a poor approximation for bubbles reaching a significant fraction of the speed of light let alone exceeding it. A full non-linear treatment of the system is necessary to produce waveforms of strong-gravity bubbles. The waveforms below are intended only as a preliminary estimate of the waveforms to be searched for. The underlying equations of weak field gravity are well established in the literature and many GR textbooks~\citep{MTW,wald1984,schutz2009,Carroll_2019}. The equation for the space--time metric perturbation $h$ propagating in the weak field limit reduces to the transverse-traceless (TT) components, which in integral form comes to
\begin{equation}
    h^{TT}_{\mu  \nu} (t,\vec{x}) \approx 4 \int d^3 y \frac{T_{\mu \nu} (t-R,\vec{y})}{R}, \label{eqn:weakperturbation}
\end{equation}
where $R = |\vec{x} - \vec{y}|$ is the flat-space displacement between source and observation points, $T_{\mu \nu}$ are the SEM tensor components, and the integral domain is over the past light cone of the observation point. In the far-field limit, where the distance $r$ from the SEM source is much larger than $Y$ the extent of the source, $r \gg Y$, the above integral can be expanded and retaining only the long-range ($\mathcal{O}(1/r)$) terms provides
\begin{align}
    &h^{TT}_{\mu  \nu} (t,\vec{x}) \approx \\ \nonumber
    &4 \int d^3 y \left[ \frac{T_{\mu \nu} (t-r,\vec{y})}{r} - \frac{\partial_0 T_{\mu \nu} (t-r,\vec{y}) \vec{x}\cdot\vec{y} }{r^2} \right] + \mathcal{O}(1/r^2),
\end{align}
where the partial derivative is with respect to the time argument of the SEM tensor. The first term of the integrand can be reduced to an expression of the energy density's time-dependent quadrupole moment, however the second term in general does not. In vacuum, $h^{TT}$  has up to two independent components referred to as the polarizations of free GWs, $h_+$ and $h_{\times}$, which can be matched directly to components of the stress tensor in Eqn.~\ref{eqn:weakperturbation}. 

\begin{figure}
\begin{center}
\includegraphics[width=1.15\columnwidth]{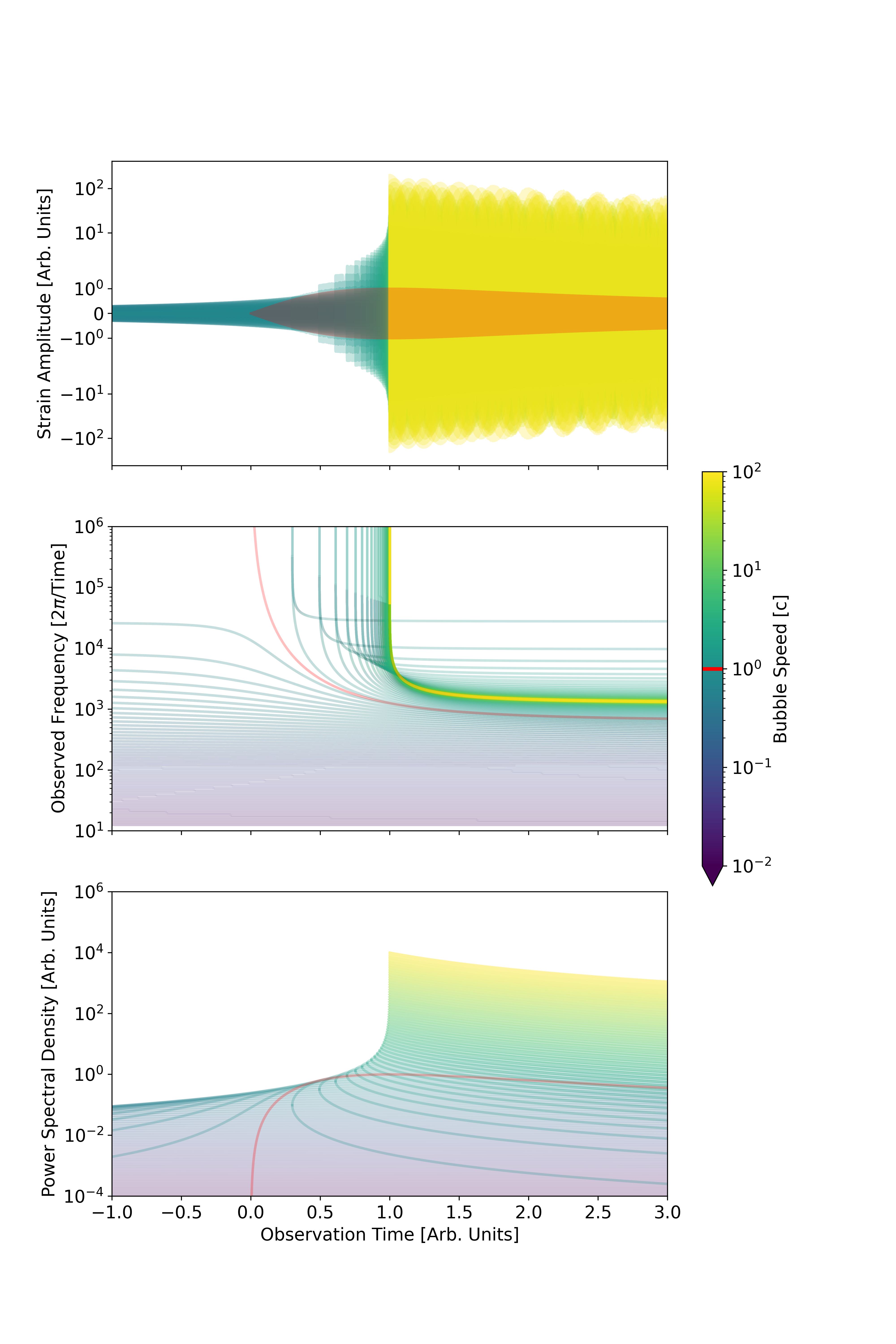}
\caption{Sample resonant gravitational waveforms measured in the $h_+$ polarization according to the basis set in Fig.~\ref{fig:diagram} and with distance of closest approach set to $d=1$. 
(Top) Time-resolved waveforms as seen at the observation point. Sub-luminal waveforms are observed at all times, while luminal and super-luminal waveforms have a time of `first light.'
(Center) Evolution of observed waveform spectra, noting that superluminal-waveforms are composed of two waves angular-ly separated viewing the bubble evolving forward in emission time (lower frequency) and backward in emission time (higher frequency).
(Bottom) Strain intensity evolution of the waveform, with separated branches for the super-luminal cases.
The light-speed craft is highlighted in red. Waveform amplitudes and spectral densities have been normalized to the maxima of the light-speed craft. }
\label{fig:GWaveforms}
\end{center}
\end{figure}

In the absence of a fully articulated geometry+SEM model of a physical warp drive we must make several prescriptive assumptions about the form of the SEM. Let us assume for the purpose of demonstration that the SEM has an oscillatory component, perhaps due to a resonance, such that the stress energy can be parameterized to repeat at regular intervals $T$ as viewed outside the bubble about its moving centroid $T_{\mu \nu} (t,\vec{y}-\vec{v}_s t) = T_{\mu \nu} (t+nT,\vec{y} - \vec{v}_s (t + nT))$ ($n \in \mathbb{Z}$), which gives the integrand a degree of regularity. For this estimate computation, let us approximate the relevant parts of the stress tensor as components parallel and perpendicular with the direction of motion of the bubble
\begin{align}
    S_{\|}(t,\vec{y}) &= T_{ij}(t,\vec{y}) \hat{v}_s^i \hat{v}_s^j = S_{\|} \delta^3(\vec{y} - \vec{v}_s t)  \cos \left( \frac{2 \pi t}{T} \right), \\
    S_{\perp}(t,\vec{y}) &= T_{ij}(t,\vec{y}) \hat{\phi}_s^i \hat{\phi}_s^j = S_{\perp} \delta^3(\vec{y} - \vec{v}_s t)  \cos \left( \frac{2 \pi t}{T} + \theta \right),
\end{align}
where $\theta$ is a constant phase offset, $\delta(\cdot)$ is a Dirac delta distribution used to localize the stress to the bubble center, and $\hat{\phi}_s$ is the craft-centered unit vector in the azimuthal direction relative to the craft's direction of motion. For some additional structure, the period of oscillation will be set inversely proportional to the speed of the craft $T = T_o c/v_s$. Some example estimated gravitational waveforms and their frequency spectra are shown in Fig.~\ref{fig:GWaveforms}.

\subsubsection{Electromagnetic Waves}

\begin{figure}
\begin{center}
\includegraphics[width=1.15\columnwidth]{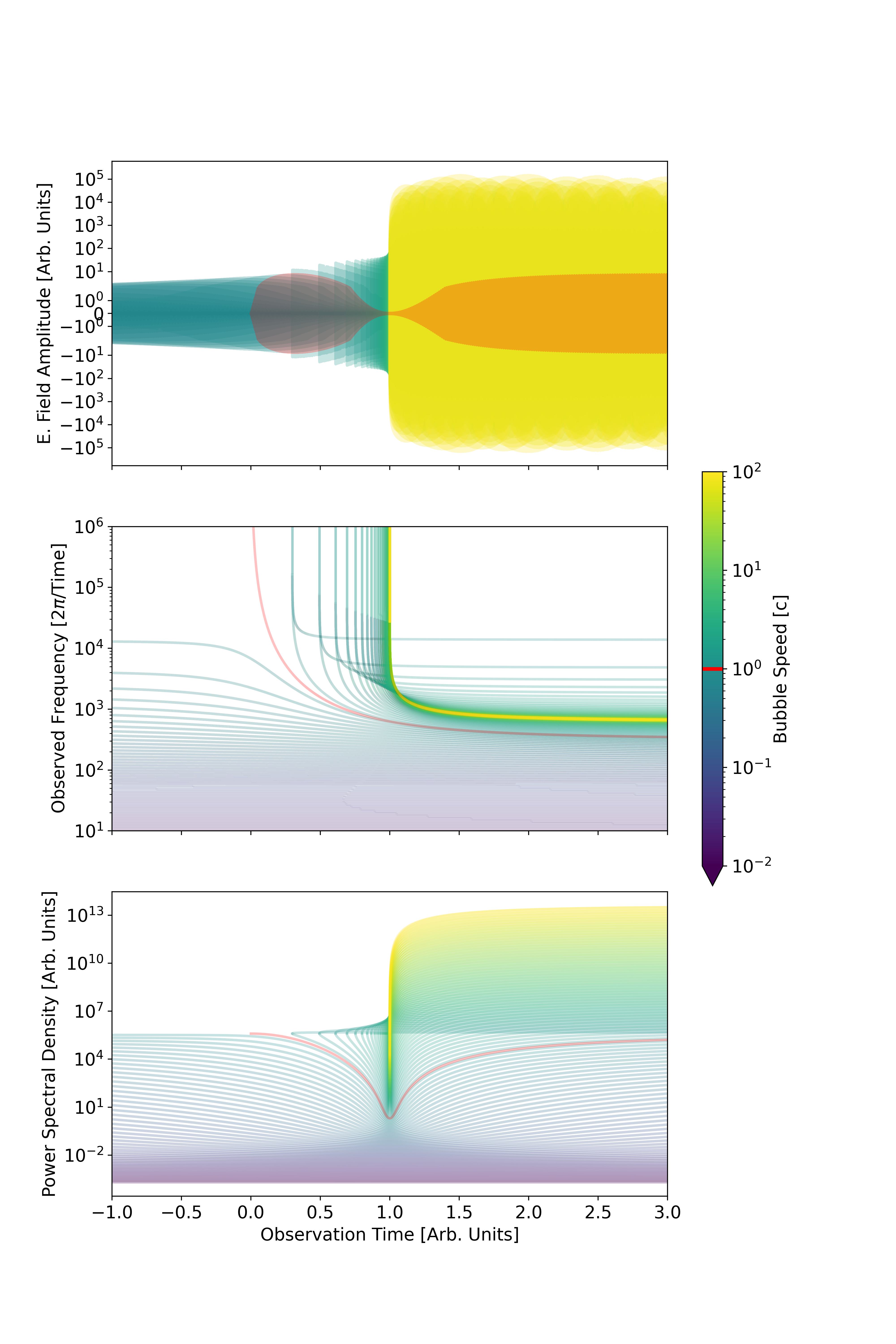}
\caption{Sample resonant electromagnetic waveforms as measured by a $\hat{z}$ antenna according to the basis set in Fig.~\ref{fig:diagram} and with distance of closest approach set to $d=1$. 
(Top) Time-resolved waveforms as seen at the observation point. Sub-luminal waveforms are observed at all times, while luminal and super-luminal waveforms have a time of `first light.'
(Center) Evolution of observed waveform spectra, noting that superluminal-waveforms are composed of two waves angular-ly separated viewing the bubble evolving forward in emission time (lower frequency) and backward in emission time (higher frequency).
(Bottom) Power intensity evolution of the waveform, with separated branches for the super-luminal cases.
The light-speed craft is highlighted in red. Waveform amplitudes and spectral densities have been normalized to the maxima of the light-speed craft.}
\label{fig:EM_waveforms}
\end{center}
\end{figure}

We next estimate the electromagnetic emissions caused by the accompanying oscillatory movement of charges in the bubble. The integral expression for the electromagnetic (spatial) vector potential in the Lorentz gauge modified from \citet{jackson_classical_1999} for curved space--times is
\begin{equation}
    \vec{A} (t,\vec{x}) = \frac{\mu_0}{4 \pi} \int d^3 y \sqrt{|g|(t-R,\vec{y})} \frac{\vec{J} (t-R,\vec{y})}{R}, \label{eqn:fullvectorfield}
\end{equation}
where $\vec{J}$ is the electric current density, $\mu_0$ is the permeability of free space, and $\sqrt{|g|}$ is the metric's contribution to the volume measure.  Similar to the GW case, the far field and weak (gravity) limit simplifies the expression to
\begin{align}
    &\vec{A} (t,\vec{x}) = \\ \nonumber
    &\frac{\mu_0}{4 \pi}  \int d^3 y \left[ \frac{\vec{J} (t-r,\vec{y})}{r} - \frac{\partial_0 \vec{J} (t-r,\vec{y}) \vec{x}\cdot\vec{y} }{r^2} \right] + \mathcal{O}(h,1/r^2).
\end{align}
The electric and magnetic fields may be computed from the vector potential in this limit via the vacuum Maxwell equations $\vec{H} = \vec{\nabla} \times \vec{A} / \mu_0$ and $ \partial_t \vec{E} = \vec{\nabla} \times \vec{H} / \epsilon_0$. The energy flux from the radiating fields is given by $\vec{S} = \vec{E} \times \vec{H}$. For this example, the movement of the currents are taken to be dominated by electric and magnetic dipoles with strengths respectively proportional to the parallel and perpendicular stresses which source the bubble, with oscillation period twice that of the stress $T_e = 2 T$. Estimated electromagnetic waveforms and their frequency spectra can be seen in Fig.~\ref{fig:EM_waveforms}. Note that while the theory of classical electromagnetism is itself linear and Eqn.~\ref{eqn:fullvectorfield} applies in the strong gravity limit when expressed covariantly, the example electromagnetic waveforms also degrade in accuracy for higher speed bubbles and the electromagnetic waveforms in this regime must also come from a full non-linear treatment of strong-gravity bubbles.

\subsubsection{High Energy Emissions}

For the emission of shorter wavelength media such as nucleons, electrons, neutrinos, higher energy photons, etc., from the sourcing SEM, we must be aware of the bubble geometry and the distribution of each species in the sourcing SEM. It is beyond the scope of this work to provide a self-consistent treatment of the system here, and we feel that the level of speculation necessary to provide a preliminary estimate for the resonance example will leave the example largely uninformative so we omit it here for later work.

\subsection{Extrinsic Emissions}
\label{sec:extrinsic}

A warp bubble `dressed' via interacting with its surroundings may produce scattering emissions. The interaction may be with compact objects which the bubble passes close to along its trajectory such as planets, stars and stellar remnants, etc. More often the bubble will encounter and pass through diffuse material such as the interstellar medium (ISM), dark matter, neutrino backgrounds, gravitational wave backgrounds, etc. This subsection concentrates on interaction with diffuse non-gravity components. The precise outcome of the scattering will depend on the details of the bubble geometry and its response to the environmental influence, but for the sake of demonstration several approximations are made: scatters will be net isotropic as viewed by the observer's rest frame; the strength of the scatter is dependent on the effective cross-sectional area of the bubble normal to its direction of motion; and the bubble geometry and therefore the scatterings will be time independent to avoid possible periodic leakage of radiation into the passenger region. Also, inspired by \cite{McMonigal_2012}, media that intersects the bubble and is scattered outwards will be given a `boost` in total energy as viewed from the observation frame amounting to
\begin{equation}
    b = 1 + \frac{\left(\vec{v}_s - \vec{v}_p\right) \cdot \hat{v}_s}{v_p},
\end{equation}
where $\vec{v}_p$ is the ISM constituent's velocity.

\begin{figure}
\begin{center}
\includegraphics[width=1.1\columnwidth]{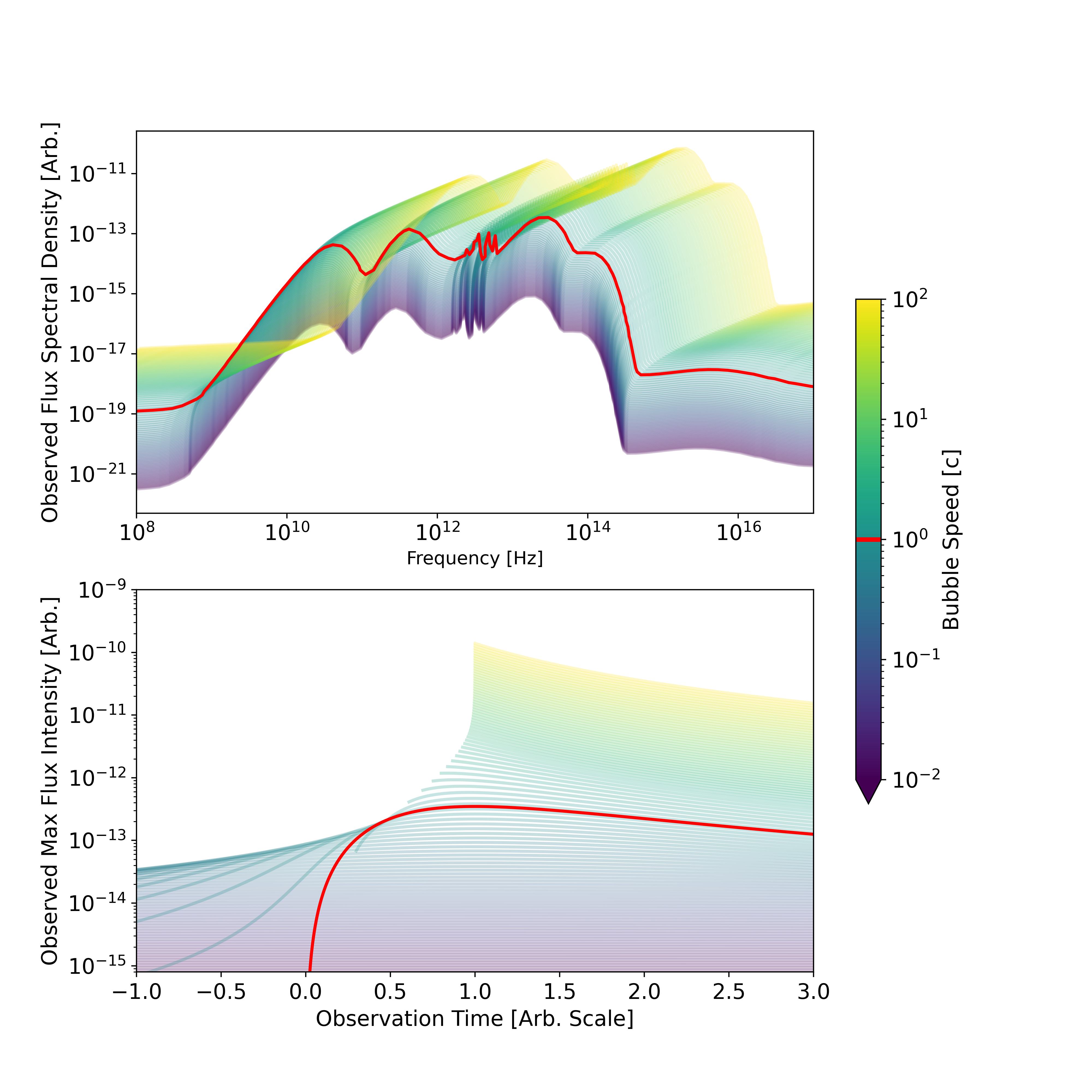}
\caption{Estimated observed emissions of ISM photons scattered by a craft moving at constant velocity. (Upper) Photon power spectrum at observation time $t=1$. (Lower) Evolution of the observed maximum flux intensity over time. The light-speed craft is highlighted in red.}
\label{fig:ISM_photon_scattering}
\end{center}
\end{figure}

The ISM consists of ionic, atomic, and molecular gases, collections of dust, cosmic rays, and electromagnetic radiation across the frequency spectrum. Electromagnetic radiation is distributed in frequency spanning from the radio to gamma ray bands, for which we will use the model given by \citep{gnedin2016physical} as representative over the sample trajectory. When disturbed by a bubble the net spectral distribution of scattered media in this simple model is of similar shape to the base spectrum but boosted in frequency and intensity, Fig.~\ref{fig:ISM_photon_scattering}.

Ambient massive particles besides neutrinos (ionic, atomic, and molecular gases, dust, etc.) will be scattered by the warp drive, but have a very limited mean free path ($\lambda_{mfp} \ll 1$~pc) and for high-speed warp bubbles will create a shock in the ISM, radiating and propagating in a way that is beyond the scope of this paper to estimate.

We limit our discussion of this signature to the overall power estimate in Section~\ref{sec:range}.

\begin{figure}
\begin{center}
\includegraphics[width=1.1\columnwidth]{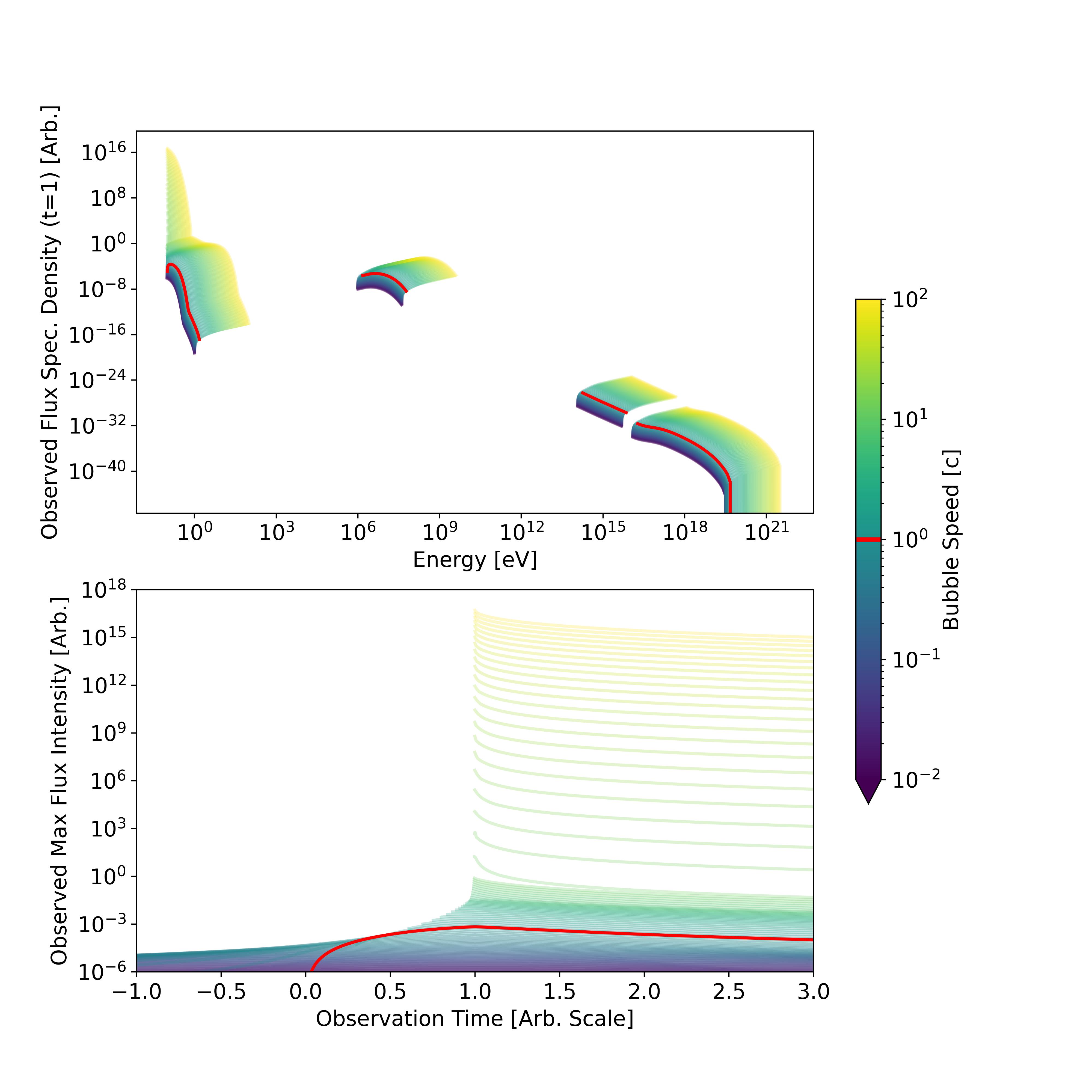}
\caption{Estimated observed emissions of a simplified ambient neutrino distribution by a craft moving at constant velocity. (Upper) Incident mass spectrum at observation time $t=1$. (Lower) Evolution of the observed maximum flux intensity over time. The light-speed craft is highlighted in red.}
\label{fig:ISM_neutrino_scattering}
\end{center}
\end{figure}

The neutrino background permits itself to a cleaner signature as it propagates relatively undisturbed after scattering. In general we should be discussing all degrees of neutrinos, whether in the mass normal basis or the flavor normal basis, but as the neutrinos' fundamental masses remain unknown we will for purposes of demonstration consider only a singular species of neutrino/anti-neutrino and project spectra to match. Estimates of the scattered neutrinos for several sub-spectra (cosmic neutrino background, diffuse supernova neutrino background, cosmogenic, and the ambient high energy neutrinos measured by Ice Cube)~\cite{neutrinospec2020} are found in Fig.~\ref{fig:ISM_neutrino_scattering} assuming a dominant neutrino mass of 0.1~eV.

There are other ambient fields expected in interstellar space including dark matter, dark energy, and the gravitational wave background, though we do not consider them here.

\subsection{Signature Intensity and Observable Range}
\label{sec:range}

The above estimates on the form of emission signatures are by themselves insufficient to determine the range of sensitivity for an observatory. For that we require two more attributes: the signature intensity and the statistical model for detection qualification of a given observatory-signature pair. For simplicity we use a signal-to-noise ratio statistic which contrasts the signature's average recorded intensity with the associated measurement uncertainty. Unique details of waveform evolution, spectral and angular shape, etc. are omitted from consideration here in favor of a plain comparison of signal intensity to random background. A threshold SNR of unity will be used as reference, with a simple relation available to scale to a more stringent level of confidence. The reference distance is set at 100~lyr between the observatory and the point of closest approach for the bubble path, with a bubble path length of 1~lyr. As the signal duration and observed intensity may vary significantly with orientation of the bubble path in addition to the speed of the bubble, we simplify by constraining the path as parameterized in Fig.~\ref{fig:diagram} with $d=100$~lyr, $L_1=L_2=0.5$~lyr. In the limit of $v_s \gg c$ the duration of the signal converges to $\Delta t = (100-\sqrt{100^2 + 0.5^2}) yr \approx 1.25\times 10^{-3}$~yrs or about 11.4~hours. The regime of bubble speed much larger than light ($v_s \gg c$) will be used for the below estimates as the average intensity scalings reduce to power laws. The length scale of the reference bubble is set to $D=1$~km, and the timescale $T_o$ of the bubble resonator will be naively set as the flat--space light crossing time $T_o = D/c$.

\subsubsection{Electromagnetic Signals}

Observatories sensitive to electromagnetic radiation are numerous and operate across a wide range of the frequency spectrum. Sources and intensity of noise also vary across the spectrum. We focus here on the radio and visible bands as preliminary estimates of signal strength and availability of archival data from ground-based observatories hint that these regions hold the best chances of producing an observation. 

Radio telescopes such as the Green Bank Telescope measure their noise spectral density in the GHz band at the level of tens of micro-Janskys to milli-Janskys depending on the target and detecting instrument~\citep{GBT_proposerguide2024}. Telescopes in the visible bands such as the NEID observatory telescopes have thermal noise levels on the order of 300~Kelvin \citep{NEID_spectrometer_noise}, or the order of $10^4$ Janskys in spectral density.

By comparison the estimated observed flux from scattered ISM photons is expected to be quite weak. The energy density of photons propagating in the ISM is on the order of $\rho_{\gamma} \sim 1$~eV/cc, implying the incident flux from scattered photons to be
\begin{align}
    &F_{EM-ISM} \sim \\ \nonumber
    &3\times10^{-7}~Jy \cdot Hz \left(\frac{v_s}{c}\right) \left(\frac{D}{km}\right)^2 \left(\frac{\rho_{\gamma}}{eV/cc}\right) \left(\frac{100~lyr}{r}\right)^2 \left(\frac{\epsilon_{\gamma}}{1.0}\right),
\end{align}
where $\epsilon_{\gamma}$ is fraction of encountered photons scattered. Note this radiation is spread across the spectrum according to the distribution given in Fig.~\ref{fig:ISM_photon_scattering}.

The energy flux from scattered ISM gas and dust components is significantly higher than that of stray photons as the energy density is measured on the order of $\rho_{m} \sim 1$~GeV/cc. If the excess energy from the scattered gas and dust interacted quickly with the surrounding ISM and re-radiated as light, the produced photon flux would be
\begin{align}
    &F_{ISM-rerad} \sim \\ \nonumber
    &3\times10^{2}~Jy \cdot Hz \left(\frac{v_s}{c}\right) \left(\frac{D}{km}\right)^2 \left(\frac{\rho_{m}}{GeV/cc}\right) \left(\frac{100~lyr}{r}\right)^2 \left(\frac{\epsilon_{m}}{1.0}\right),
\end{align}
where $\epsilon_{m}$ is fraction of encountered matter scattered. This flux would also be spread across the spectrum though we do not provide an estimate for the distribution here. 

The intensity from the example intrinsic resonant emissions are more difficult to estimate given the unknown SEM requirements on warp bubbles in practice. The estimates of \citet{Lentz2020,Bobrick2021}, and \citet{Fuchs_2024} show energy requirements in the $E_B \sim (few)\times 10^{-1} M_{\odot}$ for a fiducial bubble of 100~m radius moving at light speed. For such bubbles, a loss to EM emissions of only $0.1\%$ over a 1~lyr journey would amount to a total energy greater than the rest mass of Earth. A more economically optimistic figure of a sourcing SEM for a well-designed bubble of the fiducial light speed and size $D=1$~km would be that it has the average density of atmosphere ($10^{-3}$~g/cc) will be taken as $10^9$~kg. The estimated flux density from electromagnetic dipole losses for the economical bubble, encapsulating all dependencies on the dipole strength, emissivity, etc. in a reference loss coefficient $\epsilon_l$, are then
\begin{align}
    &f_{EM-res} \sim  \\ \nonumber
    &2\times10^6~Jy \left(\frac{v_s}{c}\right)^2 \left(\frac{100~lyr}{r}\right)^2 \left(\frac{\epsilon_l}{10^{-4}}\right) \left(\frac{M}{10^{9}~kg}\right),
\end{align}
where we have assumed a resonance width of 1~Hz. The dipole resonance will emit at an estimated center frequency of $f_e \sim 0.5 v_s / D \approx 1.3\times10^5 (v_s/c)$~Hz. The observed frequency will be much higher, in excess of two orders of magnitude, with an asymptote at the point of first light pushing the frequencies into the GHz range coincident with Green Bank. Note though  that the flux density decreases proportionally (Fig.~\ref{fig:EM_waveforms}).

\subsubsection{Gravitational Signals}

Operating gravitational wave observatories differ from electromagnetic ones in that they directly measure and quantify their uncertainty in the metric strain, not in analogous gravitational power. In the far field this translates to a slower fall off of sensitivity with distance from the source ($\sim 1/r$) compared to the inverse square scaling law of power measurements.

The uncertainty of the observatories also vary with incident wave frequency, but their region of highest sensitivity of $10^{-23}-10^{-20}$~Hz$^{-1/2}$ spans the frequency range of approximately $30-10^5$~Hz~\citep{LIGO_O3_sensitivity}.

If the losses to gravitational radiation from the internal resonance are on the same order as the resonant electromagnetic losses of $0.1\%$ for the 1~lyr journey, the local space--time strain intensity is expected to be much weaker than modern observatories' best sensitivity
\begin{align}
    &h_{GW-res} \sim \nonumber \\
    &7\times10^{-29}~Hz^{-1/2} \left(\frac{v_s}{c}\right) \left(\frac{100~lyr}{r}\right) \left(\frac{\epsilon_l}{10^{-4}}\right)^{1/2} \left(\frac{M}{10^{9}~kg}\right),
\end{align}
where again  we have assumed a resonance width of 1~Hz, and the estimate center frequency of the quadrupole resonance is $f_e \sim v_s / D \approx 2.6\times10^5 (v_s/c)$~Hz. Similar to the electromagnetic resonance, the observed frequencies of the gravitational waveform are raised by multiple orders of magnitude in this regime, placing the incident waveform out of band of current interferometric observatories.

\section{Proposed Research Plan for Warp Bubble Searches}
\label{sec:program}

The previous section named numerous possible sources of warp drive emission signatures, presenting preliminary forms of several intrinsic and extrinsic signals and estimates for their detectability range using current instruments. Unfortunately, these signal models are as-of-yet too crude to form the basis of a definitive search. This section outlines a research plan to (1) elevate the modeling and simulation of warp bubbles and their emissions to a useful level and (2) perform searches for the resulting signatures using observatories over the most promising regions of the electromagnetic spectrum.

\subsection{Improved Warp Drive Emission Models}

The increased interest in warp drives this decade has produced significant insights into the classes of solutions available using GR and the sourcing SEM that may produce physical warp drives (Section~\ref{sec:warpreview}). One such set of insights to be incorporated in this research plan is the likely need to expand beyond the Natario class of geometries to accommodate the energy conditions and positive mass theorem to include a passenger craft~\citep{Santiago2021a,Helmerich_2024,Fuchs_2024}. This increases the number of dynamical degrees in the geometry, namely the lapse function $N$ and hypersurface metric component $h_{ij}$, complicating the design of candidate space--times. Fortunately the set of tools for analyzing fully non-linear and self-consistent GR geometry+SEM and warp drives in particular has grown~\citep{xAct,Helmerich_2024} in addition to extant numerical relativity simulation codes~\citep{brandt2020einstein,GRChombo}.

The articulated geometry+SEM solutions will ideally span the three major phases of travel of a craft: creation and apparent acceleration, coast, and apparent deceleration and diffusion. While the coasting phase, the only phase for which we have provided estimated signatures, is expected to dominate the duration of a journey of significant length, the acceleration and deceleration phases are crucial to understanding how the bubble forms around a craft, establishes itself to propagate at a given velocity, and dissolves to end the journey. These phases are expected to produce the most intense and unique signals of the journey. Also, these endpoint signals are expected to occur in the vicinity of objects such as star systems and planets, highlighting regions of interest for the astronomy community, especially astrobiologists.

Not all physics and engineering challenges to warp drives must be solved in order to build models sufficient for a technosignature search. Though we wish for precise renderings of the warp drive, a simple and likely non-optimal model of geometry+SEM will be sufficient for early searches. We set out the base elements to building warp bubble emissions models as follows:

\begin{itemize}
    \item Warp drive models will be comprised of both geometric and SEM degrees and will be self-consistently simulated under full dynamical conditions
    \item Trajectories will span all phases of travel (acceleration, coast, deceleration)
    \item The bubble must be sufficiently stable to complete the journey with little risk to the passengers
    \item Journeys will be computed both in near vacuum as well as in an ISM-like environment
    \item SEM-efficient bubbles will be prioritized
\end{itemize}

Acceptable simulations will have emissions computed from the radiative leakages out of the simulation's space--time domain for each messenger type. Performing a simulation for each bubble design, size, speed, etc., to generate emissions models is inefficient. The creation and use of semi-analytic template models informed by simulations and parameterized over the most relevant inputs will allow for rapid searches over broad ranges of time and spectra.

\subsection{Technosignature Search}

\begin{figure}
\begin{center}
\includegraphics[width=\columnwidth]{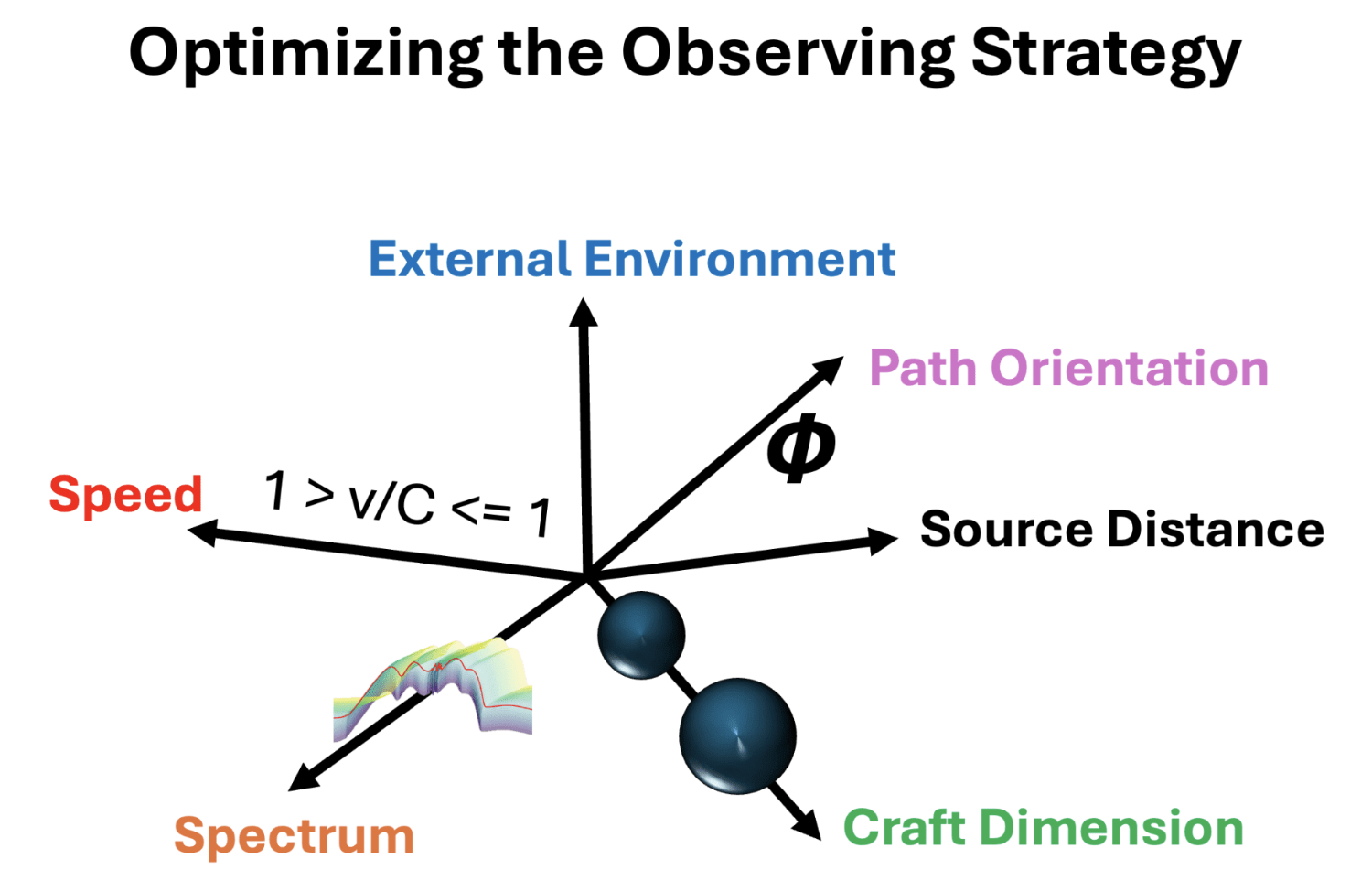}
\caption{Illustration of six key parameters and meta-parameters to be optimized over for sensitivity of a search for bubble technosignatures.}
\label{fig:optimization}
\end{center}
\end{figure}

Implementing the emissions templates into a search will begin with archived observational data then shift to observing time. To confine the scope to a manageable parameter space, we will only search for electromagnetic emission signals from stellar systems with at least one confirmed exoplanet. An optimized search strategy will maximize the likelihood of observing a signature around an active system. The leading parameters contributing to this likelihood are source distance from Earth, craft speed, orientation, and dimensions, and spectrum region (Fig.~\ref{fig:optimization}).

For our archival search we will use the publicly available Keck Observatory archive, Green Bank Observatory archive, and European Southern Observatory (ESO) archive. The Keck Observatory operates in the 0.3-5 micron range, providing us with observational data in optical and near-infrared region of the spectrum, the Green Banks Observatory operates in the radio part of the spectrum (0.1 - 116 GHz), and the ESO telescopes in the radio, optical, and infrared.

We hypothesize that intrinsic and extrinsic emissions associated with a super-luminal craft will not look the same and will appear the strongest in different regions of the EM spectrum. By using multiple archives we are able to account for these differences and improve the chances of detecting at least one of the emission types. We will also search for bi-modal features (described in Section~\ref{sec:signatures}) unique to a craft moving at super-luminal speeds. We are currently unaware of any natural phenomenon that could produce such a signal and the feature is distinct from other proposed technosignatures \citep{HAQQMISRA2022194}. However, the bi-modal  feature has a small observing time window and we expect it will be difficult to achieve a high enough signal-to-noise ratio (SNR $\thickapprox$ 100-200, \citep{tellis2015search}) to be confident of a positive detection.\\

\underline{Archival Search Campaign:}
\begin{itemize}
    \item Use Keck Observatory, Green Bank Observatory, and European Southern Observatory public archives
    \item Focus on EM spectra emissions in the optical, near-infrared, and radio
    \item Study stellar systems with at least one confirmed exoplanet
\end{itemize}

Our observing campaign will utilize the Keck Observatory, Allen Telescope Array, and possibly the Green Bank Observatory. This multi-observatory approach follows the multi-spectrum strategy from the Archival Search to improve the chances of detecting an intrinsic or extrinsic super-luminal craft emission.\\

\underline{Expanding the Search - \textit{Active Observations}}
\begin{itemize}
\item Observing with the Keck Observatory, Allen Telescope Array, and Green Banks Observatory
\item Expand on potential areas of interest from Archival Search by shifting to narrow band searches
\end{itemize}

\section{Summary}
\label{sec:closing}

In this paper we have motivated emissions from warp drives and outlined a broad research plan for their search. Preliminary estimates of the observable signatures from the operation of a coasting warp bubble in our galaxy show multiple unique characteristics such as spatial and spectral bi-modality in the superluminal case which evolves over hours, days, or many years depending on the distance, speed, and orientation of the path. We have identified emissions that may be observable beyond a range of 100~lyr using current observatories, reducing the degeneracy of signatures with naturally occurring phenomena. These emissions are also expected to occur over multiple messenger types (electromagnetic, gravitational, massive particles) revealing an opportunity to increase confidence in a signal's identity via multi-messenger corroboration. The formulated research plan seeks to undertake a near-term search for warp bubble technosignatures by (1) creating template models of emissions from physical warp drives in realistic environments over all phases of travel and (2) conducting searches for realistic signatures beginning in the electromagnetic spectrum.

\section{Acknowledgements}
\label{sec:acknowledgements}

This project was funded in part by an internal investment at Pacific Northwest National Laboratory, which is operated by Battelle for the U.S. Department of Energy. We also acknowledge consultation, through private communication, with Stephen Kane, Ravi Kopparapu, and Evan Sneed.


\bibliography{Bibliography}{}
\bibliographystyle{aasjournal}



\end{document}